\documentclass[aps,twocolumn]{revtex4}

\usepackage{graphicx}
\usepackage{psfrag}
\usepackage{fancyhdr}

\begin{document}
\date{\today}

\title{Collisionless relaxation in gravitational systems: From violent relaxation to gravothermal collapse}

\author{Yan Levin, Renato Pakter, and Felipe B. Rizzato}
\address{
  Instituto de F\'{\i}sica, UFRGS, 
  %Universidade Federal do Rio Grande do Sul \\ 
  Caixa Postal 15051, CEP 91501-970, Porto Alegre, RS, Brazil 
}  

%%%%%%%%%%%%%%%%%%%%%%%%%%%%%%%%%%%%%%%%%%%%%%%%%%%

\begin{abstract}
Theory and simulations are used to study   
collisionless relaxation of a gravitational $N$-body system.  It is shown that when the
initial one particle distribution function satisfies the virial condition -- potential
energy is minus twice the kinetic energy -- the system quickly
relaxes to a metastable state described {\it quantitatively} by the Lynden-Bell
distribution with a cutoff.  If the initial distribution function does not meet
the virial requirement, the system undergoes violent oscillations, resulting in a partial
evaporation of mass.  The leftover particles phase separate into a core-halo
structure.  The theory presented allows us to quantitatively predict 
the amount and the distribution of mass left in the central core, 
without any adjustable parameters.  On a longer time scale  $\tau_G \sim N$ 
collisionless relaxation leads to a gravothermal collapse.  

\end{abstract}


\maketitle

Since the pioneering works of Boltzmann and Gibbs, systems with long range
interactions have been a major stumbling block to the development of statistical mechanics~\cite{Pa90}.   
The difficulty was already well appreciated by Gibbs, who has
noted that the equivalence between statistical ensembles breaks down when the interparticle
potentials decay with exponents smaller than the dimensionality of the embedding space~\cite{Gib28}. 
When this happens, systems exhibit some very unusual properties which 
appear to  violate the second law of thermodynamics.
For example, confined non-neutral plasmas are found to phase separate into coexisting phases of 
{\it different} temperatures~\cite{LePa08}, while the self-gravitating
systems, such as elliptical galaxies, 
are characterized by a {\it negative} specific heat~\cite{Ly99}. 
The explanation for these counterintuitive results
lies in the fact that when the interactions are long ranged,
thermodynamic equilibrium is never reached~\cite{MiSa07} and the
laws of {\it equilibrium} thermodynamics do not apply.

In the limit in which the number of particle goes to infinity $(N \rightarrow \infty)$,
while the total mass and charge are 
kept fixed --- the so called thermodynamic limit --- the collision duration time diverges, and 
the dynamical evolution of non-neutral plasmas and gravitational systems 
is governed  {\it exactly} 
by the collisionless Boltzmann --- or as it is known in plasma physics, Vlasov 
equation~\cite{Br77}.  
This equation never reaches a stationary state --- the spatiotemporal
evolution continues  {\it ad infinitum} on smaller and smaller length scales, while the
one particle position and velocity distribution function evolves in time as an incompressible fluid. 
In practice, however, since there is always a minimum resolution
most systems do {\it appear} to evolve to a well
defined stationary state.  This state, however, is very different from the normal thermodynamic equilibrium
characterized by the Maxwell-Boltzmann distribution --- it explicitly depends on the initial
distribution of the particle positions and velocities.

Forty years ago~\cite{Ly67}, Lynden-Bell argued that although the fine-grained
distribution function of positions ${\bf r}$ and velocities ${\bf v}$, 
$f(t, {\bf r},{\bf v})$, never reaches equilibrium,    
the {\it coarse-grained}
distribution function $\bar f(t, {\bf r},{\bf v},)$, {\it averaged} on microscopic length scales,
relaxes to a meta-equilibrium with $\bar f({\bf r},{\bf v})$.   
Since in practice the very small length scales can not be resolved experimentally,  
observations and simulations can only
provide us with the information about the  coarse-grained 
distribution function $\bar f({\bf r},{\bf v})$.
To obtain $\bar f({\bf r},{\bf v})$ we divide the  
phase space into macrocells of volume ${\rm d^3{\bf r}}\,{\rm d^3{\bf v}}$, which
are in turn subdivided into $\nu$ microcells, each of volume $h^3$.  
The initial distribution function $f_0({\bf r},{\bf v})$ is discretized 
into a set of levels $\eta_j$, with $j=1...p$.  
The incompressibility of the Vlasov dynamics requires that at any time $t$ each microcell
contains at most one discretized level $\eta_j$ and that the overall hypervolume of each level
$\gamma(\eta_j)=\int \delta(f(t,{\bf r},{\bf v})-\eta_j) {\rm d^3{\bf r}}{\rm d^3{\bf v}}$, 
be preserved by the dynamics.   We denote the fraction of the volume of the {\it macrocell} 
at $({\bf r},{\bf v})$ occupied by the level $j$ 
as $\rho_j({\bf r},{\bf v})$.  Using a standard combinatorial procedure~\cite{Ly67,Ch06,LePa08} 
it is now possible to associate
a coarse-grained entropy $S$ with the distribution of $\{\rho_j\}$. 
Lynden-Bell argued that the collisionless relaxation
should lead to the density distribution of levels which is most likely, i.e. the
one that maximizes the {\it coarse-grained} entropy, consistent with the
conservation of all the dynamical 
invariants --- energy, momentum, angular momentum, and the hypervolumes $\gamma(\eta_j)$.
In terms of the volume fractions  $\{\rho_j\}$, 
the stationary distribution function becomes
$\bar f({\bf r},{\bf v})=\sum_j \eta_j \rho_j({\bf r},{\bf v})$. 
If the initial distribution has only one level $p=1$  (is water-bag),
%--------------------------- 
\begin{eqnarray}
\label{e1a} 
f_0({\bf r},{\bf v})=\eta_1 \Theta(r_m-r)\Theta(v_m-v)
\end{eqnarray}
%----------------------------- 
where $\Theta(x)$ is the Heaviside step function and  $\eta_1=9/16 \pi^2 r_m^3 v_m^3$ ---
the maximization procedure
is particularly simple, yielding a Fermi-Dirac distribution,
%--------------------------- 
\begin{eqnarray}
\label{e2}
\bar f({\bf r},{\bf v})=\eta_1 \rho({\bf r},{\bf v})=
\frac{\eta_1}{e^{\beta [\epsilon({\bf r},{\bf v})-\mu]}+1}\;.
\end{eqnarray}
%-----------------------------
In the expression above, $\epsilon$ is the mean energy of particles at position  ${\bf r}$ with velocity ${\bf v}$. 
$\beta$ and $\mu$ are two Lagrange multipliers required by the conservations of
energy and the number of particles,
%--------------------------- 
\begin{eqnarray}
\label{e3}
\int{\rm d^3{\bf r}}\,{\rm d^3{\bf v}}\;\epsilon({\bf r},{\bf v}) \bar f({\bf r},{\bf v})=\epsilon_0, \\ \nonumber 
\int{\rm d^3{\bf r}}\,{\rm d^3{\bf v}} \bar f({\bf r},{\bf v})=1\;,
\end{eqnarray}
%-----------------------------
where $\epsilon_0$ is the energy per particle of the initial distribution and the units are such that $h=1$. 
By analogy with the usual Fermi-Dirac statistics,
we define the effective temperature of a stationary state $T$ as  $\beta=1/k_B T$.
This temperature should not be confused
with the standard definition of temperature in terms of the average kinetic 
energy -- the latter  being valid only for classical systems in thermodynamic equilibrium. 
In the thermodynamic limit, the gravitational potential $\phi$ of $N$ particles with the total mass $M$ 
satisfies the Poisson equation 
%---------------------------------
\begin{eqnarray}
\label{e4}
\nabla^2 \phi = 4 \pi\, G \,m\> n({\bf r}),
\end{eqnarray}
%--------------------------------------- 
were $m=M/N$ and $n({\bf r})=N \int {\bar f}\, {\rm d}^3{\bf v}$ is the particle number density.   
The Poisson equation (\ref{e4}) and the equations (\ref{e2},\ref{e3}) form 
the basis of Lynden-Bell's violent relaxation theory~\cite{Ly67,ChSo98,AnFa07}.  
The idea is that the original distribution $f_0$ --- which is far from equilibrium i.e. is statistically
unlikely --- will  relax rapidly to $\bar f({\bf r},{\bf v})$, thus maximizing the coarse grained entropy.  
In practice, however, what is found is that self-gravitating systems usually relax to structures characterized
by dense cores surrounded by dilute halos, the distribution functions of which
are quite different from  Lynden-Bell's $\bar f$. 
The failure of the theory was attributed to the fact that the violent
relaxation occurs on very fast dynamical time scale and the system does not have time 
to explore all of the phase
space to find the most probable configurations~\cite{ArLy05}.  
Recent work on non-neutral plasmas~\cite{LePa08}, however, provides a very different picture.  
It has been found that confined non-neutral plasmas also relax to a core-halo structure.  
In that case, however, the halo production 
has been clearly shown to be the result of parametric resonances arising from the macroscopic bulk 
oscillations~\cite{Glu94}.  If the initial distribution is constructed in such a way as to suppress macroscopic
oscillations, the resulting stationary state was 
found to be precisely the one predicted by the Lynden-Bell theory~\cite{LePa08}.  It is reasonable, therefore,
to suppose that a similar mechanism will be at work for the self-gravitating systems as well.  Strong oscillations
will lead to parametric resonances --- a form of a non-linear Landau damping~\cite{Ka98} --- which will 
transfer a large amount of energy
to some particles at the expense of the rest.  These particles will either escape to infinity (evaporate) or will 
form a dilute halo which will surround the central core.

To test this theory, we first consider the case in which the
macroscopic oscillations are suppressed.  This can be achieved
by forcing the original distribution to satisfy the virial condition $2 K=-U$,   
the virial number ${\cal R} \equiv -2 K/U$ is one,  where $K$ is the total kinetic energy
and $U$ is the total gravitational energy.  To simplify the discussion, we will restrict our attention
to the initial distributions of the water-bag form $(p=1)$. For these distributions the 
virial condition reduces to the requirement that
$v_m=\sqrt{G M/r_m}$ and the average energy per particle is 
$\epsilon_0=\frac{3}{10} m v_m^2-\frac{3}{5}\frac{G M m}{r_m}$.  We expect that under these conditions  
$f_0$ will relax to the distribution given by Eq. (\ref{e2}), with $\epsilon(r,v)=m v^2/2 +m \phi(r)$, 
subject to constraints of Eqs. (\ref{e3}).
There is, however, one difficulty. Since the gravitational potential decays to zero at large distances, 
Eq.(\ref{e2}) requires that at any {\it finite} temperature there should be a 
non-vanishing particle density over {\it all} space.  The 
normalization conditions (\ref{e3}), therefore, cannot be satisfied in an infinite space.
However,  if we confine our attention
to sufficiently short times, before a significant number of particles has a chance to escape from the main 
clusters --- in practice this time is very large when the virial condition is satisfied ---  the normalization
problem can be avoided by artificially restricting the particle positions to lie within a sphere  
of radius $R$.  The
situation here is very similar to the one encountered in the theory of electrolyte solutions~\cite{Le02}.
In that case, the canonical partition function of an ionic cluster is found to diverge and 
a cutoff has to be introduced to obtain finite results.
The divergence is a natural consequence of the fact that at any {\it finite} 
temperature ionic clusters are 
unstable and will fall apart after a sufficiently long time.  On short time scales,
however, the dynamics of ionic clusters is well described by a statistical theory with a cutoff.
Furthermore, the thermodynamics of electrolyte solutions at low temperatures 
is found to be {\it completely} insensitive to the precise value of 
the cutoff used~\cite{LeFi96}.   We find the same is true for the gravitational systems as well.  
In the infinite time limit, a gravitational cluster satisfying a virial condition is  unstable
and some particles will slowly evaporate.   On ``short'' time scales, however, the cluster properties 
are well described
by a statistical theory with a cutoff.  
The precise value of the cutoff is unimportant --- as long as it is not
too large.  In our calculations we have taken the cutoff to be at $R=10 r_m$, 
but all the results remain visibly
{\it unaffected} if we replace this by $5 r_m$ or $100 r_m$.  
The cutoff-Lynden-Bell distribution (cLB)~\cite{ChSo98} 
is then given by $\bar f_{cLB}({\bf r},{\bf v})=\bar f({\bf r},{\bf v}) \Theta(R-r)$. It is also 
possible to use an energy cutoff~\cite{Ch98}, but for the purposes of the present 
calculation this is not necessary.  
We now iteratively solve the Poisson equation Eq.~(\ref{e4}) 
with the distribution  $\bar f_{cLB}$ subject to the conservation equations (\ref{e3}).
Integrating the Fermi-Dirac distribution over all velocities and taking advantage of the radial
symmetry of the distribution (\ref{e1a}) 
the Poisson's equation (\ref{e4}) takes the form
%%%%%%%%%%%%%%%%%%%%%%%%%%%%%%%%%%%%%%%
\begin{equation}
\label{e5}
{1 \over r^2} {\partial \over \partial r} r^2 
{\partial \phi \over \partial r} = - 16 G M\pi^2 \,\eta_1\,\sqrt{\pi \over 2 \tilde{\beta}^3}\,Li_{3/2}(- e^{\tilde{\beta}\left[\tilde{\mu} - \phi(r)\right]}),
\end{equation}
%%%%%%%%%%%%%%%%%%%%%%%%%%%%%%%%%%%%%%%%%%
where $\tilde{\beta}=\beta m$, $\tilde{\mu}=\mu/m$ and $Li_n(x)$ is the $n^{th}$ polylogarithm function of $x$. 
This nonlinear equation  
is solved numerically with the boundary conditions $\phi'(r=0)=0$ and $\phi(r \rightarrow \infty)=0$. 
The gravitational potential $\phi(r)$ depends parametrically on $\tilde{\mu}$ and $\tilde{\beta}$,
which are determined using the conservation equations (\ref{e3}). Integrating over
the velocities these become 
\begin{eqnarray}
\label{e6}
&-& 8 \pi^2\,\eta_1\,\sqrt{\pi \over 2}
\int_0^R \left({3 \over \tilde{\beta}^{5/2}} \, Li_{5/2}(- e^{\tilde{\beta}\left[\tilde{\mu} - \phi(r)\right]}) +\right. \nonumber \\ 
&&\left.{\phi(r) \over \tilde{\beta}^{3/2}} 
Li_{3/2}(- e^{\tilde{\beta}\left[\tilde{\mu} - \phi(r)\right]})\right)\,r^2\,dr=\varepsilon_0,\label{energy} \\
&&16 \pi^2 \eta_1 \sqrt{\pi \over 2 \tilde{\beta}^3} \int_0^R r^2 \, Li_{3/2} (- e^{\tilde{\beta}\left[\tilde{\mu} - \phi(r)\right]})
\,dr = 1 \nonumber.
\end{eqnarray}

To compare the theory with the simulations, we calculate the number 
particles inside shells located between $r$ and $r+{\rm d}r$,
$N(r){\rm d}r=4 \pi N r^2 {\rm d}r \int{\rm d^3{\bf v}} \bar f({\bf r},{\bf v})$. 
In the simulations $20,000$ particles
were initially distributed according to the water-bag distribution Eq.~(\ref{e1a}) 
and then allowed to evolve in an {\it infinite space} 
in accordance with
Newton's equations of motion.  To avoid the collisional effects and to speed up 
the simulations, the forces were calculated using the mean-field
Gauss law. As discussed earlier,
this procedure becomes exact in the thermodynamic limit. 
In Fig. 1 the solid lines are the  values of $N(r)/N$ 
obtained using the theory above, while the points are the results of the
dynamics simulation, the distances are measured in 
units of $r_m$ and the dynamical time scale is $\tau_D=\sqrt{r_m^3/G M}$.
An excellent agreement is found between the theory and the 
simulations, {\it without any adjustable parameters}. 
%%%%%%%%%%%%%%%% figure %%%%%%%%%%%%%%%%%%%%%
\begin{figure}[h]
\begin{center}
\includegraphics[width=6cm]{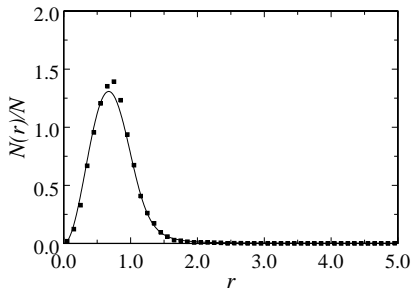}
\end{center}
%\vspace{1cm}
%\psfrag{r}{$r/\sqrt{K/\kappa_z}$}
\caption{Mass distribution for ${\cal R}=1$:  
solid curve is obtained using the cLB distribution and the points are the result of
dynamics simulation. There is no halo,  all 
mass is in the core.}
\label{fig1}
\end{figure}
%%%%%%%%%%%%% end of figure %%%%%%%%%%%%%%%%% 
To further explore the dynamics of the relaxation process, we define the temporal deviation
of the density distribution $N(r,t)$ from the stationary 
cLB value,  $N_{cLB}(r)$,
%%%%%%%%%%%%%%%%%%%%%%%%%%%%%%%%%%%%%%%
\begin{equation}
\label{e7}
\chi(t)=\frac{1}{N^2} \int_{0}^{\infty} \left[N(r,t)-N_{cLB}(r)\right]^2,
\end{equation}
%%%%%%%%%%%%%%%%%%%%%%%%%%%%%%%%%%%%%%%%%%
The inset of Fig.2a shows that after a very short time 
interval of $\tau_R \approx \tau_D $, the original distribution $f_0$ quickly relaxes 
to the cLB form. Furthermore the relaxation time $\tau_R$  
is independent on the number of 
particles in the system --- this is precisely the Lynden-Bell's violent
relaxation regime.  The 
metastable cLB distribution persists 
until a finite fraction of particles evaporates from the main cluster. 
Following the violent
relaxation, $\chi$ begins to increase again.  The rate
of this increase depends strongly on the number of particles in the system, Fig 2a.  
We define $\tau_G$ as the time at which the (violently) relaxed
distribution  begins to  deviate from the cLB form by $1\%$, $\chi=0.01$.   
This time depends on $N$ as $\tau_G \approx 4 N \tau_D$.  Scaling the simulation time 
with $\tau_G$ gives
an excellent collapse of the data on a single universal curve, see Fig. 2b.
%%%%%%%%%%%%%%%% figure %%%%%%%%%%%%%%%%%%%%%
\begin{figure}[h]
\begin{center}
\includegraphics[width=7cm]{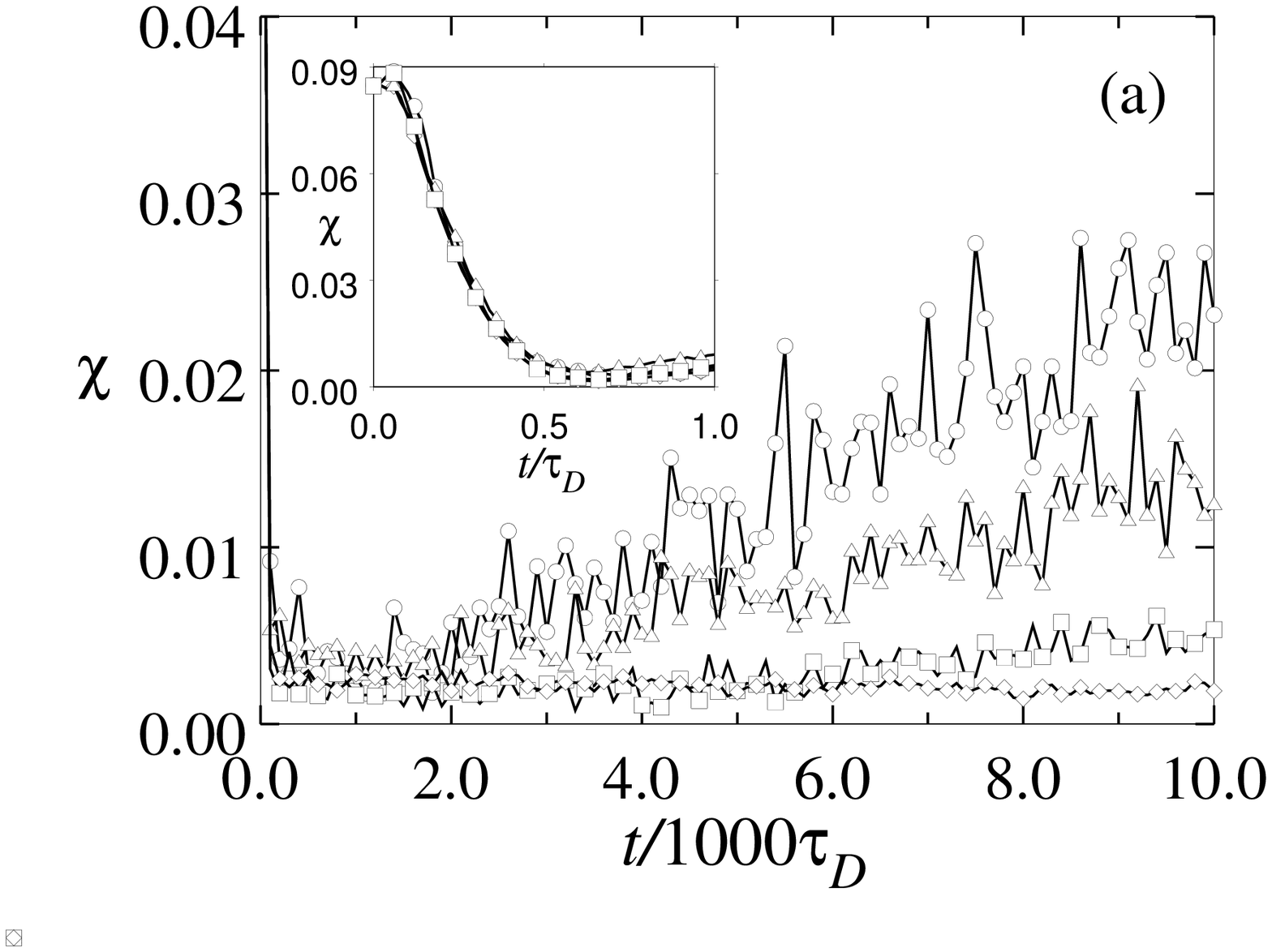}
\includegraphics[width=7cm]{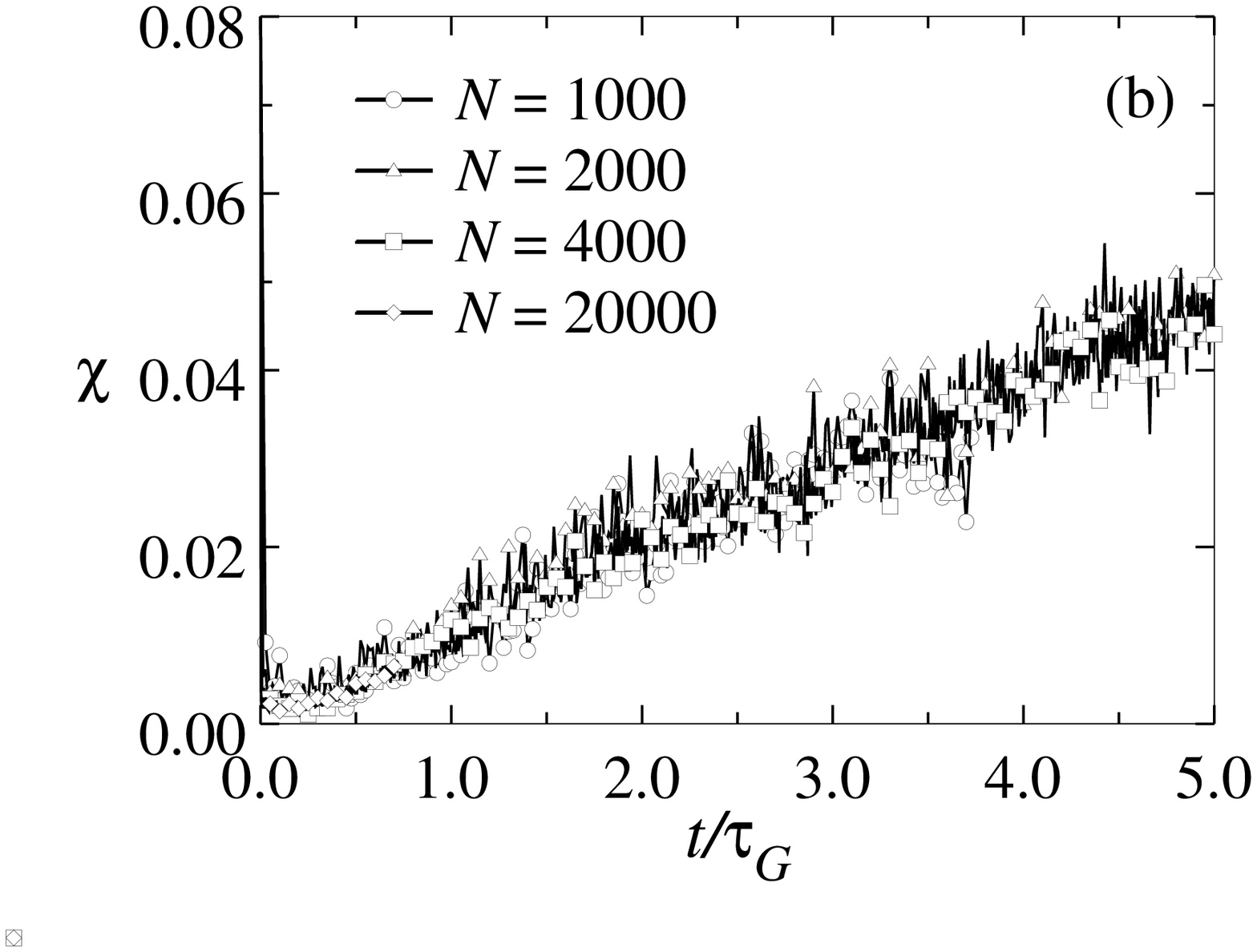}
\end{center}
%\vspace{1cm}
%\psfrag{r}{$r/\sqrt{K/\kappa_z}$}
\caption{(a) $\chi(t)$ for different number of particles in the system.  
Inset shows the violent relaxation regime which occurs on a very short
time scale $\tau_D$ and is independent of $N$.  
Following the violent relaxation, the system undergoes small oscillations which die out after
about $100\tau_D$. At this time the distribution is  precisely of the cLB form.  On a longer
time scale $\tau_G$ the system undergoes a gravothermal collapse.  The rate of this
collapse is a  linear function of $N$. (b) When the time is scaled with
$\tau_G$, all the data in (a) collapses on one universal curve. }
\label{fig2}
\end{figure}
%%%%%%%%%%%%% end of figure %%%%%%%%%%%%%%%%% 
The fact that  $\tau_G$ diverges with $N$ implies that in the thermodynamic limit the  cLB distribution 
will last forever!
We conclude that if the virial condition is satisfied and the macroscopic oscillations are suppressed, 
the phase-mixing --- linear Landau damping~\cite{Sa93} ---  mechanism is 
extremely efficient to produce a local
ergodicity.   For times larger than $\tau_G$, the phase space incompressibility condition
$\rho_1({\bf r}, {\bf v}) \le 1$ is violated and the system undergoes a slow gravothermal
collapse.  We note, however, that since the time scale $\tau_G$, is larger than
the  Chandrasekhar time $\tau_{Ch} \sim  \tau_D N/\ln(N) $, the binary collisions omitted in our
simulations must be explicitly taken into account to study this regime.

Small deviations from the virial condition $ 0.8<{\cal R}<1.2$ 
result only in weak oscillation which are not sufficient to produce  
significant parametric resonances. 
Thus, we find that for this range of virial numbers the 
cLB theory remains in  good agreement with the dynamics simulations. 
For larger deviations from  ${\cal R}=1$, 
the situation changes dramatically.  In these cases, 
the initial distribution 
undergoes violent oscillations resulting in a partial mass evaporation 
and a halo production.  
A fraction of the particles quickly gain enough energy from the resonances to completely escape from the
main cluster (evaporate), while the other fraction gains only enough energy to   
move away from the core, remaining gravitationally bound to it.  
This latter class  forms a dilute halo surrounding the dense central core. 
The evaporation and halo production progressively cool down the 
cluster until all the collective
oscillation cease at $T=0$. The particles
left in the core should then be in the energy ground state, with their 
distribution function given by that of a fully degenerate Fermi gas 
$\bar f_{c}({\bf r},{\bf v})=\eta_1\,\Theta(\mu-\epsilon)$.  
This is precisely what is found when the cutoff 
$R$  in the cLB distribution is extended to infinity.  In this limit the cLB distribution splits into
two domains --- a compact zero temperature core described by  $\bar f_{c}({\bf r},{\bf v})$ 
plus an evaporated fraction of zero energy particles at infinity. Integrating over velocities,
the Poisson equation becomes
%%%%%%%%%%%%%%%%%%%%%%%%%%%%
\begin{equation}
{1 \over r^2} {\partial \over \partial r} r^2 
{\partial \phi \over \partial r}= {32 G M \pi^2 \sqrt{2} \eta_1 \over 3} (\tilde{\mu} - \phi(r))^{3/2}\>\Theta(\tilde{\mu} - \phi(r)), 
\end{equation}
%%%%%%%%%%%%%%%%%%%%%%%%%%%%
where the value of $\tilde{\mu}$ is determined by the energy conservation,
%%%%%%%%%%%%%%%%%%%%%%%%%%%%%%%%
\begin{equation}
- {32 \pi^2 \sqrt{2} \eta_1\over 5}  
\int_0^\infty r^2 (\tilde{\mu} - \phi(r))^{5/2} \> 
\Theta(\tilde{\mu} - \phi(r)) \, dr =\varepsilon_0\,.
\end{equation}
%%%%%%%%%%%%%%%%%%%%%%%%%%%%%%%%%%%%%%%%%%%%%%%%
The norm of $\bar f_{c}$ then gives the amount of mass left in the central core after 
the process of collisionless relaxation
is completed. Fig.4 shows that the theoretically predicted $\bar f_{c}$
is in excellent agreement with the core mass distribution obtained using the dynamics simulations. 
However, in order to have a complete account of the halo mass distribution
a more detailed dynamical study is necessary.  The work in this direction is now in progress. 
On a time scale larger than $\tau_G$ the core is, once again, 
found to undergo a gravothermal collapse.
%%%%%%%%%%%%%%%% figure %%%%%%%%%%%%%%%%%%%%%
\begin{figure}[h]
\begin{center}
\includegraphics[width=6cm]{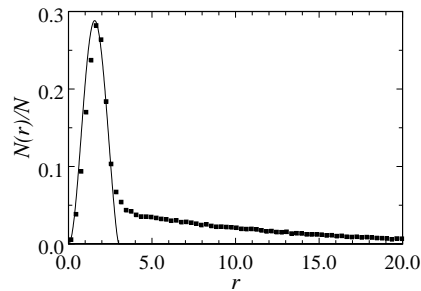}
\end{center}
%\vspace{1cm}
%\psfrag{r}{$r/\sqrt{K/\kappa_z}$}
\caption{ Mass distribution for ${\cal R}=1.9$ -- points are the
result of the simulation.  Solid curve obtained using $\bar f_{c}$ gives the 
mass distribution inside the core. The theory predicts that $54\%$ of the mass will 
be in the halo or will evaporate, which is in agreement with the simulations. 
There are no adjustable parameters}
\label{fig3}
\end{figure}
%%%%%%%%%%%%% end of figure %%%%%%%%%%%%%%%%%    

Traditionally the failure of the Lynden-Bell's theory was attributed to the 
fact that for gravitational systems relaxation occurs on a short dynamical 
time scale $\tau_D$ so that the system has no time
to explore all of the phase space to find the most ``probable'' configuration. 
Present work, however, provides a very different picture.   
Strong oscillations lead to propagating density waves~\cite{Mo98} and to parametric resonances 
which force some particles
into statistically improbable regions of the phase space.   
These regions do not mix with the rest of the system.
When the oscillations (parametric resonances) are suppressed, the mixing is very efficient and the 
predictions of the Lynden-Bell theory are verified quantitatively.  
Outside the virial condition, strong oscillations lead to a partial 
mass evaporation and a core-halo coexistence.
The theory presented here gives a quantitatively accurate account of the core, it also allows
us to predict the amount of mass which will be lost to the halo production and evaporation.

This work is partially supported by CNPq and by the 
US-AFOSR under the grant FA9550-06-1-0345.

%\bibliographystyle{prsty}
%\bibliography{references}

\begin{thebibliography}{99} 

%\bibitem{Kl67} M. J. Klein, Science {\bf 157}, 509 (1967).

\bibitem{Pa90} T. Padmanabhan, Physics Reports {\bf 188}, 285 (1990).

\bibitem{Gib28} J. W. Gibbs, {\it Collected Works}, Longmans, Green and Co.,
NY (1928).

\bibitem{LePa08} Y. Levin, R. Pakter, and T. N. Telles, Phys. Rev. Lett. {\bf 100}, 040604 (2008).

\bibitem{Ly99} D. Lynden-Bell  and R.M. Lynden-Bell, Mon.
Not. R. Astro. Soc. {\bf 181}, 405 (1977); D. Lynden-Bell, Physica A {\bf 263}, 293 (1999);
%D.J.Wales and R.S.Berry, Phys. Rev. Lett. {\bf 73}, 2875 (1994);  
%R.M.Lynden-Bell and D.J. Wales, J. Chem. Phys. {bf 101}, 1460 (1994); 

\bibitem{MiSa07} K. Michaelian and I. Santamaria-Holek, EPL {\bf 79}, 43001 (2007); 
D. Lynden-Bell and R.M. Lynden-Bell, EPL {\bf 82}, 43001 (2008); 
K. Michaelian and I. Santamaria-Holek, EPL {\bf 82}, 43002 (2008)

\bibitem{Br77} W. Braun and K. Hepp, Comm. Math. Phys.  {\bf 56}, 101 (1977);
A. Antoniazzi,1 F. Califano, D. Fanelli, and S. Ruffo, Phys. Rev. Lett. {\bf 98}. 150602 (2007). 

\bibitem{Ly67} D. Lynden-Bell, Mon. Not. R. Astron. Soc. {\bf 136}, 101 (1967).

\bibitem{Ch06} P.-H. Chavanis, Physica A {\bf 359}, 177 (2006).

\bibitem{ChSo98} P.-H. Chavanis and J. Sommeria, Mon. Not. R. Astron. Soc. {\bf 296}, 569 (1998).

\bibitem{AnFa07} A. Antoniazzi,1 D. Fanelli, J. Barr\', P.-H. Chavanis,  T. Dauxois, and S. Ruffo,
Phys. Rev. E {\bf 75}, 011112 (2007). 



\bibitem{ArLy05} I. Arad and D. Lynden-Bell,  Mon. Not. R. Astron. Soc. {\bf 361}, 385 (2005)

\bibitem{Glu94} R.L. Gluckstern, Phys. Rev. Lett. {\bf 73}, 1247 (1994); 
R. P. Nunes, R. Pakter, and F. B. Rizzato, Phys. Plasmas, {\bf 14}, 023104 (2007).

\bibitem{Ka98} H.E. Kandrup, Annals New York Acad. Sci. {\bf 848}, 28 (1998). 

\bibitem{Le02} Y. Levin, Rep. Prog. Phys. {\bf 65}, 1577 (2002).

\bibitem{LeFi96} Y. Levin and M.E. Fisher, Physica A {\bf 225}, 164 (1996).

\bibitem{Ch98} P.-H. Chavanis,  Mon. Not. R. Astron. Soc. {\bf 300}, 981 (1998)

\bibitem{Sa93} D. Sagan, Am. J. Phys.  {\bf 62}, 450  (1993). 

\bibitem{Mo98} P.J. Morrison, Rev. Mod. Phys. {\bf 70}, 467 (1998); 
F.B. Rizzato, R. Pakter, and Y. Levin, Phys. Plasmas {\bf 14}, 110701 (2007).

%\bibitem{NaFrWh96} J.F. Navarro, C.S. Frenk, and S.D.M. White, Astrophys. J. {\bf 462}, 563 (1996).


\end{thebibliography}
%\input{ref.bbl}

\end{document}